\documentclass[preprint2]{aastex}
\usepackage{epsfig}

\def\beq{\begin{eqnarray}}
\def\eeq{\end{eqnarray}}

\def\rvir{R_{\rm vir}} 
\def\Vvir{V_{\rm vir}}
\def\vrms{v_{\rm rms}}
\def\cvir{c_{\rm vir}}
\def\lp{\lambda'}
\def\sgl{\sigma_{\lambda}}
\def\sg{\sigma_g}
\def\omm{\Omega_{\rm m}}
\def\oml{\Omega_{\Lambda}}

\title{Modeling Angular-Momentum History in Dark-Matter Halos}	 

\author{Ariyeh H. Maller, Avishai Dekel} 
\affil{Racah Institute for Physics, The Hebrew University,  
Jerusalem 91904, Israel} 
\and 
\author{Rachel S. Somerville} 
\affil{Institute of Astronomy, Madingley Road, Cambridge CB3 0HA, UK} 
 
\begin{document} 
 
\begin{abstract} 
We model the acquisition of spin by dark-matter halos in semi-analytic 
merger trees.  We explore two different algorithms; one in which halo 
spin is acquired from the orbital angular momentum of merging 
satellites, and another in which halo spin is gained via tidal 
torquing on shells of material while still in the linear regime.  We 
find that both scenarios produce the characteristic spin distribution 
of halos found in N-body simulations, namely, a log-normal distribution 
with mean $\approx 0.04$ and standard deviation $\approx 0.5$ in the log.
A perfect match requires fine-tuning of two free parameters.
Both algorithms also reproduce the general
insensitivity of the spin distribution 
to halo mass, redshift and cosmology seen in N-body simulations. 
The spin distribution can be made strictly constant 
by physically motivated scalings of the free parameters. 
In addition, both schemes predict that halos which have had recent major
mergers have systematically larger spin values.  These algorithms can 
be implemented within semi-analytic models of galaxy formation based 
on merger trees.  They yield detailed predictions of galaxy properties 
that strongly depend on angular momentum (such as size and surface 
brightness) as a function of merger history and environment. 

\end{abstract} 
 
\keywords{galaxies:formation--galaxies:spiral} 
 
%%%%%%%%%%%%%%%%%%%%%%%%%%%%%%%%%%%%%%%%%%%%%%%%%%%%%%% 
\section{Introduction} 
\label{sec:intro} 
 
Understanding the origin of spin in rotationally supported disk 
galaxies is clearly a crucial part of any theory of galaxy formation 
\citep{hoyle:51}. The general theory \citep{fe:80} reproduces galactic 
disks with roughly correct sizes, based on the assumptions that the 
gas originally had the same specific angular momentum as dark-matter 
(DM) halos do today, and that the infalling gas conserves specific 
angular momentum \citep{mest:63}. 
 
The origin of angular momentum in dark-matter halos can be understood 
in terms of linear tidal torque theory in which protohalos are torqued 
by the surrounding shear field 
\citep*{hoyle:51,peeb:69,doro:70,white:84,be:87,sb:95,pdh:01}. Many 
models for the formation of galactic disks have been proposed based on 
the general picture of \citeauthor{fe:80}. Some of these models 
construct disk galaxies solely from the properties of the final halos 
\citep*{blum:86,flor:93,dss:97,mmw:98,bosch:00}, and others incorporate 
the mass-accretion histories of the halos 
\citep*{wf:91,kwg:93,cole:94,cole:00,kauf:96,afh:98,fa:00,sp:99,spf:01}. 
 
The halo spin is a crucial element for modeling important physical 
quantities such as galactic-disk size, surface brightness, star 
formation rate, and rotation velocity in semi-analytic models. 
However, in none of the previous models is the angular momentum of DM 
halos assigned self-consistently based on their mass accretion and 
merger histories, primarily because an appropriate simple recipe for 
this buildup of spin has not yet been developed. The aim of this paper 
is to provide such a recipe. 
 
We explore two scenarios.  In the first, halo spin is generated by the 
transfer of orbital angular momentum from satellites that merge with 
the halo (orbital-merger scenario). This scenario was suggested by 
\citet{vitv:01}, who also compare it with simulations in more 
detail than is done here. In the second, linear 
tidal-torque theory is applied to spherical shells of infalling matter 
(tidal-torque scenario).  Each recipe is applied in turn, within Monte 
Carlo realizations of halo merger histories generated using the 
extended Press-Schechter approximation and the method of 
\citet{sk:99}. We test and calibrate these scenarios by comparing 
their predictions for the distribution of spin with the known results 
of N-body simulations. 
 
The outline of the paper is as follows. Section~\ref{sec:back} 
presents our notation and defines the quantities of 
interest. Section~\ref{sec_orb} explores the orbital-merger scenario, 
while Section~\ref{sec_ttt} addresses the tidal-torque picture.  In 
\S~\ref{sec_ind} we discuss the dependence of the spin distribution on 
mass, cosmology and redshift.  In \S~\ref{sec_mhd} we study the 
dependence of the spin distribution on halo merger history. In 
\S~\ref{sec_conc} we conclude and provide a brief discussion 
concerning the implications for disk sizes. 
 
%%%%%%%%%%%%%%%%%%%%%%%%%%%%%%%%%%%%%%%%%%%%%%%%%%%%%%%%%%%%%%%%%%% 
\section{Background} 
\label{sec:back} 
The angular momentum of a halo, $J$, is commonly expressed in terms of 
the dimensionless spin parameter \citep{peeb:69} 
\begin{eqnarray} 
\lambda=J\sqrt{|E|}/GM^{5/2},  
\end{eqnarray}  
where $E$ is the internal energy of the halo.  In practice, the 
computation or measurement of this quantity, especially the energy, 
may be ambiguous.  Furthermore, it introduces an undesired dependence 
on the specifics of the halo density profile, which depends on 
cosmology, redshift and halo mass \citep[see][]{bull:01} and may also 
depend on merging history \citep{wech:01a}.  Instead, following 
\citet{bull:01b}, we will focus on the distribution of the variable 
\begin{eqnarray} 
\lp={{J}\over{\sqrt{2}M\Vvir\rvir}} ,  
\end{eqnarray}  
which is more straightforward to compute and does not explicitly 
depend on the density profile of the halo.  For a singular isothermal 
sphere truncated at a radius $\rvir$, this parameter is equal to the 
traditional spin parameter, $\lp=\lambda$. They are also approximately 
equal for halos with a more realistic density profile: an NFW profile 
\citep{nfw:96} with concentration $\cvir\approx 10$. 
 
Many studies based on N-body simulations have revealed that the 
distribution of spin parameter $\lambda$ is well fit by a log-normal 
distribution, and varies very little as a function of halo mass, 
redshift, or cosmological model. \citet{bull:01b} have demonstrated 
that the distribution of $\lp$ is also well fit by a log-normal 
distribution, namely, 
\begin{eqnarray}  
P(\lp)d\lp={{1}\over{\sqrt{2\pi\sgl^2}}} 
\exp{\left(-{{\ln^2{(\lp/\lp_0)}}\over{2\sgl^2}}\right)} 
{{d\lp}\over{\lp}} , 
\label{eq:lognormal} 
\end{eqnarray}  
with $\lp_0=0.035$ and $\sgl=0.5$\footnote{ Note that in the 
log-normal distribution $\lp_0$ is the spin corresponding to the mean 
of $\ln(\lp)$, and $\sgl$ is the standard deviation of $\ln(\lp)$. We 
sometimes refer loosely to $\lp_0$ and $\sgl$ as the mean and standard 
deviation of the log-normal distribution.} 
 
\section{The Orbital Merger Scenario} 
\label{sec_orb} 
 
In this section we explore the assumption that all of the orbital 
angular momentum of a merger is converted into the spin of the merger 
product. For simplicity we will consider all mergers as involving two 
bodies and refer to the more massive of the two progenitors as the 
halo and the other as the satellite.  Following the merger tree 
algorithm \citep{sk:99}, we neglect any mass or spin loss.  We can 
safely neglect the internal spin of the incoming satellite because 
even for equal mass mergers the spin is typically only $\sim 10\%$ of 
the orbital angular momentum. 
 
%-------------------------------- 
\subsection{Encounter Parameters} 
 
Consider a satellite whose center is at a distance $r$ from the center 
of the halo and moving with a velocity $\vec{v}$ relative to the halo. 
Then in center of mass coordinates, the energy and angular momentum 
are given by 
\begin{eqnarray}  
E_{orb}&=& \frac{1}{2}\mu v^2-G{{M_{h}m_{s}}\over{r}}\\  
L_{orb}&=& \mu \vec{v} \times \vec{r} 
\end{eqnarray}  
where $h$ and $s$ denote the halo and satellite respectively and $\mu$ 
is the reduced mass, $\mu=M_hm_s/(M_h+m_s)$.  We will assume that the 
satellite comes from outside of the halo and therefore $r$ is at least 
as large as the virial radius of the halo ($r \geq \rvir$). 
For equal mass mergers ($M_h=m_s$) this implies that at some moment 
we can express the orbital energy in terms of the virial velocity of 
the halo 
\begin{eqnarray} 
E_{orb}= m_s\left({{v^2}\over{4}}-\Vvir^2\right),  
\end{eqnarray} 
which makes it clear that for a given energy there is a maximum 
orbital angular momentum $L_{orb} \leq \mu v \rvir$ for the 
halo-satellite system.  While this particular expression is only valid 
for equal mass mergers, the situation is similar for other encounters. 
 
\begin{figure} [t]%fig1 
\centering 
\vspace{15pt} 
\epsfig{file=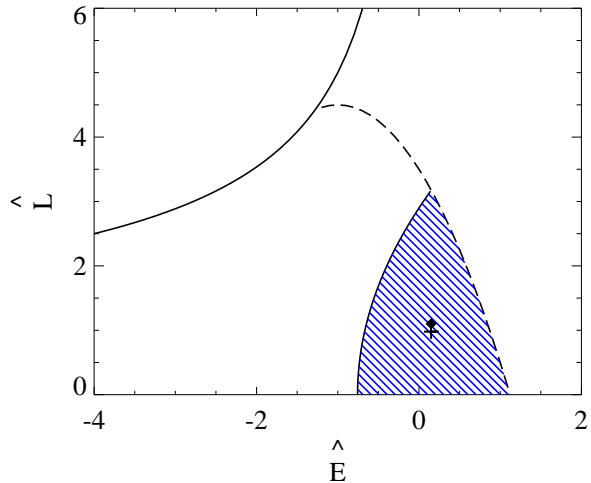,width=\linewidth} 
\vspace{10pt} 
\caption{Initial encounter parameters.  The parameter space of orbit 
energy ($\hat{E}$) and angular momentum ($\hat{L}$) for equal mass 
halos, following \protect\citet{bt:87}.  The region above the thick 
solid line is excluded by definition.  The region below the dashed 
line refers to encounters that will result in mergers within a Hubble 
time.  The region below the thin solid line refers to halos that 
are separated by at least $\rvir$.  The cross marks the average value 
of $\hat{L}$, and the diamond marks the value we use in our model.} 
\label{fig_EL} 
\end{figure} 
 
\citet{bt:87} introduce dimensionless energy and angular momentum 
variables (Equation 7-85):  
\begin{eqnarray} 
\hat{E}={{E_{orb}}\over{\frac{1}{2}\mu \vrms^2}} \; ;\;  
\hat{L}={{L_{orb}}\over{\mu r_{e}\vrms}}  
\end{eqnarray}  
where $r_{e}$ and $\vrms$ are the half mass radius and rms velocity of 
the halo. For a halo with the NFW density profile and a concentration 
of $\approx15$, $r_{e}=0.3\rvir$ and $\vrms=1.2\Vvir$.  For equal mass 
non-rotating spherical halos, the possible orbital parameters are 
shown in Figure \ref{fig_EL}.  The dashed line excludes encounters 
that take longer then a Hubble time to merge.  Adding the additional 
constraint that the satellite starts entirely outside of the halo, 
(i.e. $r \geq \rvir$) leaves the hatched region as the permissible range 
of encounter parameters.  Figure \ref{fig_EL} is only valid for equal 
mass mergers; in general the possible values of $\hat{E}$ and 
$\hat{L}$ will depend on $M_h$ and $m_s$. 
 
Since we do not know the probability distributions of $\hat{E}$ or 
$\hat{L}$, we introduce a fudge factor $f$ with which to parameterize 
our ignorance. Thus, we will take the orbital angular momentum of the 
satellite-halo system to be given by 
\begin{eqnarray} 
L_{orb}= f \mu \Vvir r. 
\end{eqnarray} 
If all permissible values for $E_{orb}$ are equally likely, we would 
deduce from Fig. \ref{fig_EL} that a typical value for $\hat{L}$ is 
about 1, corresponding to $f \approx 0.34$.  In principle, $f$ could 
depend on mass and/or redshift but we will take it to be a constant 
and adjust it to get a good fit to the observed distribution of spin 
from N-body simulations; below we will find that $f=0.38$ yields the 
best results. It would also be possible to determine the value of $f$ 
empirically by analyzing N-body simulations \citep{vitv:01}. 
 
We generate mass accretion histories based on the formalism of 
\citet{sk:99} with the slight modification introduced by 
\citet*{bkw:00}.  We produce 500 random realizations of the merger 
history of a halo, adopting a fixed value of $f$. The direction of the 
orbital angular momentum, $\theta$ (specifically, the angle between 
the orbital angular momentum vector of the satellite and the spin 
vector of the halo), is drawn at random from a uniform or Gaussian 
distribution. For the uniform case, we choose $\gamma=\cos(\theta)$ 
from a uniform distribution between -1 and 1. 
For the Gaussian distribution, we choose values of $\gamma$ between
-1 and 1 from:  
\begin{eqnarray} 
P(\gamma) {\rm d}\gamma \propto 
\exp{\left(-{{(\gamma-1)^2}\over{2\gamma_m^2}}\right)}
{\rm d}\cos\gamma \, ,  
\label{eq:gauss} 
\end{eqnarray}  
where we set the free parameter $\gamma_m$ (the width of the 
distribution of directions) empirically. We repeat this 10 times for 
each history, resulting in a total of 5000 realizations. Unless 
specified otherwise we consider a halo at $z=0$ of mass $5 \times 
10^{11} M_{\sun}$ in a $\Lambda$CDM cosmology with $\omm=0.3$, 
$\oml=0.7$, $H_0=70 \;$kms$^{-1}$Mpc$^{-1}$, and $\sigma_8=1.0$. 
 
%------------------------------------------- 
\subsection{Results} 
 
\begin{figure} [t]%fig2 
\centering 
\vspace{15pt} 
\epsfig{file=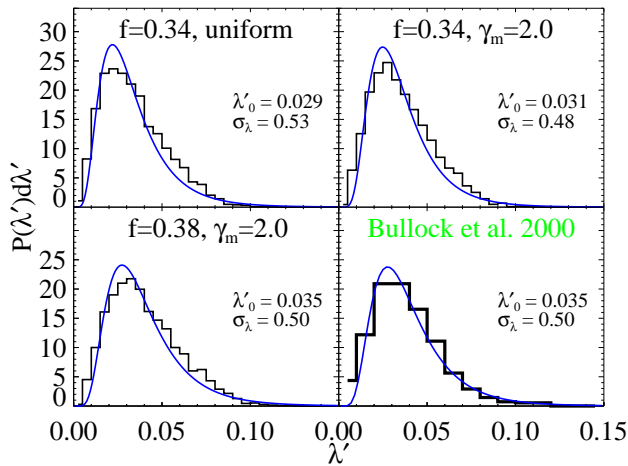,width=\linewidth} 
\vspace{10pt} 
\caption{ The distribution of $\lp$ predicted by the orbital merger 
scenario for different values of $f$ and $\gamma_m$, compared to the 
simulation data of \protect\citet{bull:01b} shown in the bottom-right 
panel. A choice of $\gamma_m=2.$ and $f=0.38$ results in a 
$\lp$ distribution identical to that seen in the simulations.  } 
\label{fig_lorb} 
\end{figure} 
 
The top-left panel of Figure \ref{fig_lorb} shows the resulting 
distribution of $\lp$ for the choice $f=0.34$ and a uniform 
distribution of directions $\theta$.  We note immediately that the 
distribution of $\lp$ is well fit by a log-normal distribution, 
equation~\ref{eq:lognormal}, with values of $\lp_0$ and $\sgl$ fairly 
close to those seen in N-body simulations, shown for reference in the 
bottom-right panel. 
 
We note that the value of $\sgl$ obtained with a fixed $f$ and a 
uniform distribution of directions is somewhat larger then that seen 
in the simulations.  This can be cured if we relax the assumption of a 
uniform distribution of directions, and allow for some correlation 
between the directions of successive mergers, as detected in N-body 
simulations on scales of a few hundreds of kpc \citep{deke:00}. We 
approximate this correlation by drawing the direction angle $\theta$ 
from a Gaussian (eqn.~\ref{eq:gauss}, above).  As seen in the 
upper-right panel of Fig. \ref{fig_lorb}, the choice 
$\gamma_m=2.0$ reduces the value of $\sgl$.  The desired value of 
$\lp_0$ and $\sgl$ are now obtained by choosing a somewhat higher
value of $f=0.38$ (Fig. \ref{fig_lorb}, bottom-left panel). 
 
In reality, the value of $f$ will also vary from encounter to 
encounter, resulting in a distribution with some scatter. This would 
tend to increase the scatter in $\lp$, which can be compensated for by 
a somewhat stronger correlation between the directions, namely a 
further decrease in the value of $\gamma_m$. For example if $f$ is 
normally distributed with a standard deviation of $0.1$, then to match 
the $\lp$ distribution requires $\gamma_m=1.0$ and the mean 
value of $f$ to be reduced to $0.32$. Here, as we don't know the form 
of the distribution of $f$ we will keep things simple by assuming it 
to have one value.  Later, our recipe could be refined by extracting 
from N-body simulations the detailed distributions of $f$ and 
$\theta$, and their possible dependence on mass and redshift. 
 
%%%%%%%%%%%%%%%%%%%%%%%%%%%%%%%%%%%%%%%%%%%%% 
\section{The Tidal Torque Scenario}  
\label{sec_ttt} 
 
The standard picture of the origin of the angular momentum of DM 
halos, as originally introduced by \citet{peeb:69} and then further 
developed by \citet{doro:70} and \citet{white:84}, is formulated in 
the context of linear tidal torque theory. Here, material acquires its 
angular momentum from the large-scale tidal field while still in the 
linear regime. We now develop an algorithm for the generation of angular
momentum in virialized halos based on the tidal 
torque picture and a spherical collapse model. 
 
Linear theory \citep[e.g.][]{white:84} predicts that the angular 
momentum of material before turn-around, at time $t$, is given by: 
\begin{eqnarray} 
J_i(t) = a(t)^2 \dot D(t)\, \epsilon_{ijk}\, T_{jl}\, I_{lk} \ , 
\label{eq:ttt} 
\end{eqnarray} 
where the time growth is from some fiducial initial time $t_{\rm i}$, 
$I_{lk}$ is the inertia tensor of the protohalo at $t_{\rm i}$, and 
$T_{jl}$ is the tidal (or shear) tensor at the halo center, smoothed 
on the halo scale.  This is based on assuming the Zel'dovich 
approximation for the velocities inside the protohalo, and a 2$^{\rm 
nd}$-order Taylor expansion of the potential.  \citet{pd:01} have 
shown that the implied standard scaling relation should be slightly 
modified. When applied to a collapsed shell of mass $m$ and comoving 
radius $q$, it reads 
\begin{eqnarray} 
{\rm d}J=  g\, a^2(t_c)\dot{D}(t_c)\, \sigma(M_h)\, m\, q^2 
\end{eqnarray} 
where $\sigma(M_h)$ is the rms density fluctuation on scale of the 
mass interior to the shell at $t_{\rm i}$.   
The time $t_c$ is when the shell practically stopped gaining spin.  An 
empirical fit from simulations shows that the effective time is, on 
average, about one third of the spherical-model collapse time, namely 
slightly before turn-around \citep*{pdhb:01}.  The parameter $g$ 
represents the small mis-alignment between the inertia and tidal 
tensors and we expect it to be $\sim 0.1$ \citep{pdh:01}. 
 
We incorporate this into the merger trees by identifying the mass in a 
collapsing shell with the total mass of merging satellites (progenitor 
halos) and accreted mass during the same period of time.  Of course 
the value of $g$ will vary from halo to halo, so we assume that $g$ is 
distributed normally with a mean $g_0$ and a standard deviation $\sg$. 
Based on the smoothness of the tidal field across the protohalo 
volume, we assume that the angular momentum of each shell is in the 
same direction and can be added linearly.  Since N-body simulations 
find that there is some nonlinear evolution of the spin direction at 
late epochs \citep{pdh:01}, we anticipate that $g$ will be forced to 
vary with $z$.
 
We again produce 5000 values of $\lp$ by considering 500 mass 
accretion histories with 10 randomly chosen values of $g$.  At $z=0$ 
in a $\Lambda$CDM cosmology, and for a halo mass of $5 \times 10^{11} 
M_{\sun}$, we find that the values $g_0=0.125$ and $\sg=0.44g_0$ yield 
a good fit to the $\lp$ distribution in simulations (log normal with 
$\lp_0=0.35$ and $\sgl=0.50$). 
 
%%%%%%%%%%%%%%%%%%%%%%%%%%%%%%%%%%%%%%%%%%%%%%%%%%%%%%%%%%% 
\section{Independence of halo mass, cosmology and redshift} 
\label{sec_ind} 
  
An interesting property of the distribution of the halo spin 
parameter, $\lambda$, as measured in cosmological N-body simulations, 
is that it seems to be independent of halo mass, redshift and 
cosmology \citep{lk:99}.  
Any model of angular momentum acquisition should therefore produce the 
same spin distribution irrespective of these three parameters. 
Indeed, our simple models seems to show this ubiquity of the $\lambda$ 
distribution of halos, or can be adjusted to reproduce this 
independence by simple, physically motivated adjustments of the free 
parameters. 
 
\begin{figure} [t]%fig3 
\centering 
\vspace{15pt} 
\epsfig{file=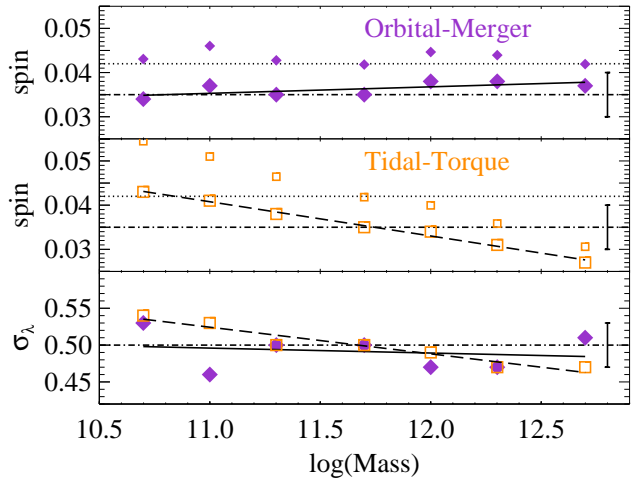,width=\linewidth} 
\vspace{10pt} 
\caption{The mean and scatter of the spin at $z=0$ as a function of 
halo mass, as derived from the two algorithms.  The upper panel shows 
$\lp_0$ (large connected symbols) and the corresponding $\lambda_0$ 
(small symbols) for the orbital-merger picture with $f=0.38$ and 
$\gamma_m=2.0$.  The middle panel is the analog for the 
tidal-torque picture with $g_0=0.125$ and $\sg=0.44g_0$.  The bottom 
panel shows $\sgl$ for the two scenarios.  The horizontal dot-dashed 
lines show the mass-independent values found in 
\protect\citet{bull:01}, with the corresponding errors as error bars. 
The dotted line is the mean value of $\lambda$ from 
\protect\citeauthor{bull:01}.  } 
\label{fig_lmorb} 
\end{figure} 
 
%-------------------------------------- 
\subsection{Halo Mass} 
 
Figure \ref{fig_lmorb} shows the values of $\lp_0$ as a 
function of halo mass with the orbital-merger picture (upper panel, 
large diamonds) and the tidal-torque scenario (middle panel, large 
squares).  Also $\sgl$ as a function of halo mass is shown for 
both scenarios (bottom panel).  
We see that over the mass range most relevant for spiral 
galaxies ($10^{10}-10^{12} M_{\sun}$) the distributions are all 
consistent with each other within the estimated errors (and the 
differences between different simulations). 

For the orbital-merger 
model there is a slight apparent trend for $\lp_0$ to increase and for 
$\sgl$ to decrease as the mass increases.  Note that when moving from 
$\lp$ to $\lambda$ this trend is expected to weaken because the halo 
concentration parameter is also a function of halo mass.  Using the
trend of average $\cvir$ with halo mass found in \citep{bull:01b}
we can convert each $\lp$ value to a value of $\lambda$, though this 
ignores the spread found in values of the $\cvir$ and any 
possible correlations between $\lp$ and $\cvir$.  The 
corresponding average values of $\lambda$ are also plotted (upper 
panel, small diamonds) in Fig. \ref{fig_lmorb} and show the absence of 
a trend with mass.   
In the simulations, there are some reports of an apparent trend of 
$\lambda$ with halo mass, in the opposite direction \citep{be:87}, but 
this trend is weak and not seen in more recent simulations. 

For the tidal-torque method different mass halos as seen in 
Figure \ref{fig_lmorb} are all in rough  
agreement with the N-body results, though there is a trend towards 
a decrease in $\lp_0$ and $\sgl$ as mass increases.   
The trend in $\lp_0$ is in the opposite direction to that found  
in the orbital-merger scenario discussed above (which means that  
the dependence of concentration on mass implies that the trend for 
$\lambda$ is even stronger (middle panel, small squares)). 

Thus both scenarios produce values of $\lp$ consistent with the results
of simulations, though both seem to show some weak mass dependence.  
Whether this weak mass dependence can also be found in simulations remains
to be seen.  Since the trends with mass of the two scenarios are in
opposite directions, detection of such a trend would tend to favor one
scenario over the other.  

Both trends can be removed by scaling one
of the free parameters with mass. The functional forms 
\begin{eqnarray}
\label{eq:scal1}
f(M_h,z) = &0.35 & \left({{M_h}\over{M_*}}\right)^{-0.02}
\end{eqnarray}
\begin{eqnarray}
 g_0(M_h,z) = &0.17& \left({{M_h}\over{M_*}}\right)^{0.08},
\label{eq:scal2}
\end{eqnarray}
where $M_*$ is the characteristic nonlinear mass \citep[see][]{lc:93},
create $\lp$ distributions with no mass dependence (note that
$\sg$ remains $0.44g_0$). Our reason for scaling these expressions in 
terms of $M_*$, which is a function of redshift and cosmology, will 
become apparent below (especially \S\ref{sub_sec_red}).

For the orbital merger picture the weak correction 
would imply that the incoming angular momentum of satellites decreases
slightly as a the halo's mass increases.  
In the tidal-torque picture we find that to remove the trend of $\lp$
with mass requires that the misalignment between the tidal 
tensor and the protohalo inertia tensor is larger for more massive halos.
Both of these conditions can be checked in N-body simulations. 

We also observe a slight trend of $\sgl$ to decrease with increasing 
mass in both scenarios.  This is a natural outcome of our modeling where
the spread in $\lp$ values arises from the spread in halo formation 
histories.   It is unlikely for a halo to have a progenitor at a 
given redshift with a mass much greater then the value of $M_*$ at that
redshift.  Thus the more massive a halo is the less likely it is to 
have assembled most of its mass long ago.  This naturally leads 
to a narrower range of possible histories for more massive halos, 
and thus to the observed trend of $\sgl$ with halo mass. The effect is 
weaker in the orbital-merger algorithm where the spread in $\lp$ is
also caused by the random orientations of the incoming satellites.
Future studies of simulations will be able to reveal whether such a 
trend exists. 
 
\begin{figure} [t]%fig4 
\centering 
\vspace{15pt} 
\epsfig{file=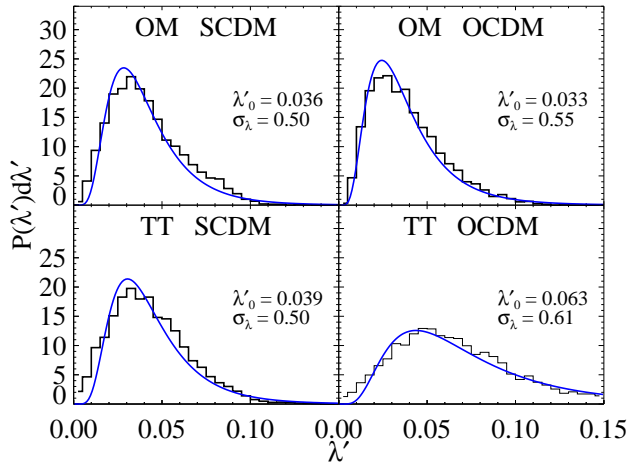,width=\linewidth} 
\vspace{10pt} 
\caption{The distribution of $\lp$ in SCDM and OCDM cosmologies 
predicted by the orbital-merger (OM) and tidal-torque (TT) models. 
}\label{fig_cos} 
\end{figure} 
 
%-------------------- 
\subsection{Cosmology} 
 
To test the dependence on cosmology, we generate merger trees based on 
two other cosmological models: the original SCDM model, with an 
Einstein-deSitter cosmology ($\omm=1.0,\oml=0.0, H_0=50 \;{\rm 
kms}^{-1}{\rm Mpc}^{-1}, \sigma_8=0.5$), and an OCDM model, 
with an open cosmology ($\omm=0.3,\oml=0.0, 
H_0=70 \;{\rm kms}^{-1}{\rm Mpc}^{-1}, \sigma_8=1.0$). 
We compute the predicted $\lp$ distribution in each of these 
cosmological models using the same fiducial values of the parameters 
$f$, $\gamma_m$, $g_0$ and $\sg$ that produced the best fit to the 
$\lp$ distribution in the $\Lambda$CDM cosmology.  The predictions of 
the orbital-merger scenario are found to be consistent with the 
distribution of $\lp$ in simulations both for SCDM and OCDM.  The 
tidal-torque scenario recovers the distribution for the SCDM 
cosmology, but it predicts a somewhat higher value of $\lp_0$ for 
OCDM.  These distributions slightly improve if the scaling with 
$M_*$ proposed in the previous section is used. Of course the 
distribution of $\lp$ in other cosmologies can be made identical 
to that of the $\Lambda$CDM cosmology by choosing slightly different 
free parameters.  Again it will require higher resolution simulations
to see if such trends exist or not.
 
%---------------------- 
\subsection{Redshift} \label{sub_sec_red} 
\begin{figure} [t]%fig4 
\centering 
\vspace{15pt} 
\epsfig{file=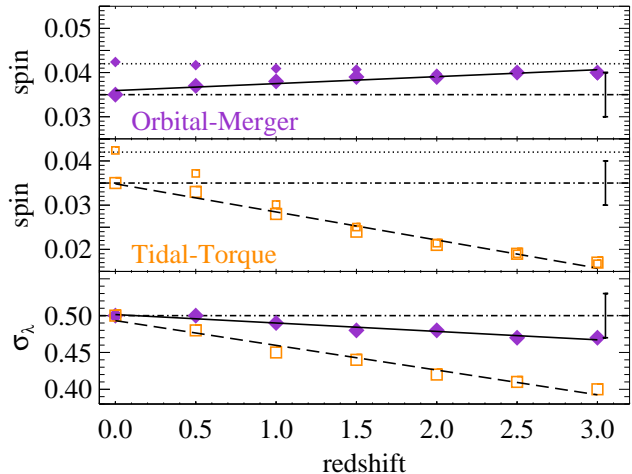,width=\linewidth} 
\vspace{10pt} 
\caption{ The mean and standard deviation of the spin parameters for 
halos of mass $5\times 10^{11}M_{\sun}$ as a function of redshift, as 
derived by the two algorithms.  The notation is as in 
Figure~\ref{fig_lmorb}. Note that by redshift 1.5 concentrations 
become so low that $\lambda$ values overlap the $\lp$ values.} 
\label{fig_lzorb} 
\end{figure}

Figure \ref{fig_lzorb} shows the $\lp$ distribution as predicted  
by the two scenarios across the redshift range 0 to 3. 
We use the $\Lambda$CDM cosmology and the fiducial 
fixed values for the  parameters $f$, $\gamma_m$, $g_0$ and $\sg$.  
 
In the orbital-merger case there is a slight increase of the mean 
value of $\lp$ with redshift. As in the case of differing masses, the 
change in the average value of the standard spin parameter $\lambda$ is
weaker because concentrations evolve with redshift as $(1+z)^{-1}$. 
The mean value of $\lambda$ is also shown in Fig. \ref{fig_lzorb}, and 
shows almost no trend with redshift. It will be interesting to see 
whether the trend in the average value of $\lp$ is verified in N-body 
simulations. 
 
The tidal-torque method with fixed parameters produces a noticeable 
evolution in the mean value of $\lp$, dominated by the fact that 
shells that collapse later have had more time to gain spin via tidal 
torques.  The evolution is even stronger when $\lambda$ is considered 
because of the evolution of $\cvir$ with redshift mentioned above.  In 
order to keep the predicted $\lp$ distribution constant with redshift, 
one needs to vary the parameter $g_0$.  
Using the scaling of the free parameters $f$ and $g_0$ introduced
above in terms of $M_h/M_*$ (equations \ref{eq:scal1} and 
\ref{eq:scal2}), the redshift dependence of the $\lp$ distribution 
is removed.

In both scenarios we see a trend for $\sgl$ to decrease with 
increasing redshift.  Considering the mass of the halo over $M_*$ 
it is apparent that this is the same effect as seen with variation 
in halo mass; halos with masses closer to or greater then 
$M_*$ have less of a spread in formation history which leads to a 
smaller spread in their $\lp$ distribution.

%%%%%%%%%%%%%%%%%%%%%%%%%%%%%%%%%%%%%%%%%%%%%%%%%% 
\section{Merger History Dependence} \label{sec_mhd} 
 
\begin{figure} [t]%fig5 
\centering 
\vspace{15pt} 
\epsfig{file=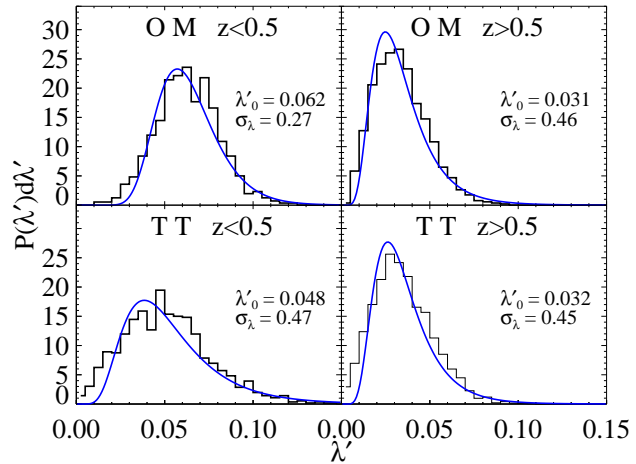,width=\linewidth} 
\vspace{10pt} 
\caption{Correlation of spin and merger history.  The distribution of 
$\lp$ for halos that have and have not undergone major mergers since 
z=0.5 in both the orbital-merger (OM, top panels) and tidal-torque 
(TT, bottom panels) pictures is shown.  It is surprising how similar 
the distributions are in these rather different pictures.  We note 
that the log-normal distribution is not as good a fit to those halos 
that are recent major-merger remnants.} 
\label{fig_lmzorb} 
\end{figure}

Having shown that the $\lambda$ distribution of halos resulting from 
our schemes can be made independent of cosmology, redshift or mass, in 
agreement with what is found in N-body simulations, we now turn to 
where our scheme does predict a $\lambda$ dependence --- the merger 
history.  Preliminary results from N-body simulations suggest 
\citep{gard:00,wech:01b} that the distribution of $\lambda$ is 
different for halos that have recently undergone a major merger. 
Using the same definition for a major merger (progenitor mass ratio of 
1:3), we see a similar effect in our models.  Figure \ref{fig_lmzorb} 
shows a significant difference between the distribution and mean value 
of $\lp$ for halos that have undergone a major merger since $z=0.5$ in 
both scenarios.  In work in progress based on a $\Lambda$CDM 
simulation, \citeauthor{wech:01b} find a mean value of 
$\log{\lp}=0.044$ for halos identified at $z=0$ that have had a major 
merger at $z < 0.5$.  Those that have not had a major merger since 
$z=0.5$ have a log average of $0.032$, and those that have not had 
such a merger since $z=2$ have a log average of $0.030$. 
\citeauthor{gard:00} finds a similar result, a log average lambda 
value of $0.049$ for halos that have experienced major mergers since 
$z=0.5$ in an OCDM model.\footnote{ A direct comparison between this 
simulation and our method is difficult because of differences in time 
step, halo identification, the concentration of halos, and the 
cosmological model.  These factors make it unsurprising that we find 
1120 out of 5000 halos to be recent major mergers while 
\citeauthor{gard:00} finds only 15 out of 1609. Our results for the 
fraction of halos with recent mergers are in general agreement with the 
simulations analyzed by \cite{wech:01b}, where a more consistent 
comparison is possible.} 
 
Our two different scenarios predict similar values for the mean in 
the log of the $\lp$ distribution for halos with recent major mergers, 
0.062 and 0.048 for the orbital-merger and tidal-torque scenarios 
respectively. The tidal-torque value is in good agreement with the 
simulations while the orbital-merger scenario gives a result 
$\approx 30\%$ higher then that found in simulations. If there is a 
situation where this algorithm might 
be suspect it would be the case of nearly equal mass mergers with high 
orbital angular momentum.  In this situation our 
assumption that we can neglect mass loss is probably invalid, 
and if mass loss occurs there will inevitably also be associated 
loss of angular momentum. 
However, recent major mergers are commonly held to result in 
ellipticals so the effect of our overestimating $\lp$ for these halos
will have a negligible effect on disk galaxies.
Our approach could be refined by treating the expected mass and 
angular momentum loss in a scheme similar to the one we have proposed. 
 
Mergers seem to leave a discernible mark on the halo's spin all the 
way back to $z=2$.  Table 1 shows the values of $\lp_0$ and $\sgl$ for 
the two scenarios for halos that have and have not undergone major 
mergers since redshifts of 0.5, 1.0 and 2.0.\footnote{ Like 
\citeauthor{wech:01b} and unlike \citeauthor{gard:00}, we find that 
the $\lp$ of non-mergers is different than the population as a whole. 
This is likely because we identify a significant fraction of halos as 
having undergone major merger events.} This strong connection between 
the merger and mass accretion history of a dark matter halo and its 
spin clearly has many interesting implications for galaxy formation 
models.  This will be best investigated in detail using full 
semi-analytic models, but we discuss qualitatively some of the 
implications below. 
 
\begin{table*} 
%\caption{Spin Distribution Dependence on Merger History} 
\begin{center} 
\begin{tabular}{lcccccccccc} 
\tableline 
\phm{z=1.0} & \multicolumn{5}{c}{Mergers}   
& \multicolumn{5}{c}{Nonmergers} \cr 
\multicolumn{2}{c}{\phm{something}} & \multicolumn{2}{c}{orbital merger} & 
\multicolumn{2}{c}{tidal torque} & \phm{x halos} & 
\multicolumn{2}{c}{orbital merger} &\multicolumn{2}{c}{tidal torque} \cr 
\phm{z=1.0} & fraction of halos & $\lp_0$ & $\sgl$ & $\lp_0$ & $\sgl$ &  
fraction of halos & $\lp_0$ & $\sgl$ & $\lp_0$ & $\sgl$ \cr 
\tableline 
\tableline 
z = 0.0 & \nodata & \nodata & \nodata & \nodata & \nodata 
& 1.0 & 0.035 & 0.50 & 0.035 & 0.50 \cr 
$z<0.5$ & 0.22  & 0.062 & 0.27 & 0.048 & 0.47 &  
0.78 & 0.031 & 0.46 & 0.032 & 0.45 \cr 
$z<1.0$ & 0.43 & 0.053 & 0.33 & 0.044 & 0.47 &  
0.57 & 0.027 & 0.41 & 0.032 & 0.45 \cr 
$z<2.0$ & 0.68 & 0.044 & 0.41 & 0.039 & 0.48 &  
0.32 & 0.023 & 0.41 & 0.028 & 0.44 \cr 
\tableline 
\end{tabular} 
\tablecomments{The table lists value of $\lp_0$ and $\sgl$ for 
those halos that have suffered a major merger (3:1)  
since redshifts 0.5, 1.0 and 2.0 and those that have not. Also  
listed are the number of halos that fall into each category.} 
\end{center} 
\end{table*} 
 
%%%%%%%%%%%%%%%%%%%%%%%%%%%%%%%%%%%%%%% 
\section{Discussion} \label{sec_conc} 
 
We have presented algorithms for tracing the acquisition of spin 
angular momentum by dark matter halos through their mass accretion or 
merger history. We have demonstrated that both methods reproduce the 
distribution of spin parameter found in N-body simulations when a 
small number of physically motivated free parameters are tuned 
appropriately. 

We have proposed two such schemes, one based on the transfer of the 
orbital angular momentum of merging satellites to the internal spin of 
the halo, and another based on tidal-torque theory applied to 
spherical shells of collapsing material. It is interesting that these 
two limits of the linear and non-linear regimes produce similar 
results. In principle, the tidal field around a halo should influence 
both the halo's accretion history and the orbital angular momentum of 
satellites, so that in the orbital-merger model, the tidal field is 
still ultimately responsible for the generation of angular momentum. 
If we knew the details of the density field around the halo then we 
could calculate the orientation and magnitude of the angular momenta 
of the merging satellites instead of drawing them randomly in a 
somewhat arbitrary fashion, which is of course what an N-body 
simulation does. 

We obtain spin distributions with respect to halo mass, 
redshift and differing cosmologies that are consistent within the 
uncertainties found in N-body simulations.  We also find weak trends
of the $\lp$ distribution with mass and redshift that are in the opposite
directions for the two scenarios.  The dependence of concentration on
mass and redshift found by \citet{bull:01b} creates
a $\lambda$ distribution without a mass or redshift trend in the orbital-
merger scenario, but increases the systematic variation of $\lp$ with 
redshift for the tidal-torque picture.  The discovery of such a trend 
in simulations would be a strong conformation of our modeling and 
its direction could be used to discriminate between the two scenarios.

We also find a trend for $\sgl$ to decrease as mass or redshift increases.
This can be understood in terms of the spread in formation histories of 
the dark matter halo.  The more massive a halo is in terms of $M_h/M_*$
the less likely it is to have been in place for a long time as that would
require progenitors with masses larger then $M_*$ 
of the corresponding time. The smaller spread 
in formation histories in our modeling translates directly into a smaller
spread in values of $\lp$.  This also explains the trend of $\lp_0$ 
with mass and redshift found in the orbital-merger picture.  As $M_h/M_*$
increases the prevalence of recent major mergers increases which as we
have shown in \S \ref{sec_mhd} correlates with higher spin values.

This effect should also be seen in the tidal-torque method; however, there
are two other effects in this case that go in the opposite direction. 
Recall from equation \ref{eq:ttt} that the angular momentum gained by 
each shell is proportional to $\sigma(M_h)$. For larger masses 
$\sigma(M_h)$ is smaller and this leads to lower spin values.  Likewise, 
at higher redshifts shells have had less time to gain angular momentum.  
These effects dominate over the expected increase in $\lp$ because of 
late time accretion. 

Because we are unsure if such trends exist in N-body simulations,
we also include simple scaling relations in $M_h/M_*$ for one of the 
free parameters in each scenario ($f$ or $g_0$) that effectively 
removes the trends with mass and redshift.  N-body simulations can be 
used to test whether $g_0$ and $f$ are functions of $M_h/M_*$.

We have obtained the interesting prediction, supported by preliminary 
recent work with N-body simulations, that there is a strong correlation 
between spin value and merger history. In particular, the distribution 
of spins for halos that have experienced recent major mergers is 
skewed towards higher values. This effect is discernible for major 
mergers at least as far back in time as $z \sim 2$.  Further study of 
N-body simulations will test how accurately our algorithms trace this 
correlation between spin and merger history \citep{wech:01b}.
  
\cite{vitv:01} use N-body simulations to measure the distribution of 
the free parameter we term $f$ and find its most common value to be
$\approx 0.5-0.6$. 
While this is higher then the value of $f$ we have used ($0.38$), they 
also find that to fit the $\lambda$ distribution they need to assume 
that $25\%$ of the angular momentum is lost during the merger event, 
implying that the effective value of $f$ is very close to what we have 
found. Closer study of simulations will be required before we 
understand the details of how orbital angular momentum is transfered 
to spin in mergers and what dependencies are involved; however, the 
simple prescription we have provided here should be adequate for many 
applications. 
 
These recipes can now be incorporated in semi-analytic methods that 
attempt to follow all the relevant physical processes of gas cooling 
and collapse, star formation, supernovae feedback, etc, over the 
formation history of galaxies.  The self-consistent treatment of 
angular momentum acquisition that we have proposed is a significant 
improvement over previous work, in which spins were assigned to halos 
at random, from the overall distribution of $\lambda$ irrespective of 
the halo's merger history.  
 
We can already anticipate some of the consequences of the predicted 
strong dependence of spin on merger history.  As an immediate example, 
this may explain the small spread in disk sizes found by 
\citet{dl:00}.  They argue that the standard \citet{fe:80} picture of 
disk formation leads to a relation $R_d \varpropto \lambda 
L_I^{\beta}$ between the disk size, the spin parameter and the near-IR 
luminosity, implying that the spread in disk size at a given 
luminosity is log-normal with standard deviation $\simeq \sgl$. 
However, they deduce from an observed sample of late-type spirals 
$\sgl=0.36\pm0.03$, which is many ``sigma'' away from the N-body 
result of $\simeq 0.5$. However, Table 1 demonstrates that considering 
subsamples of the halo population, divided by merger history, results 
in lower values of $\sgl$.  If we assume that late-type galaxies 
inhabit halos that have not experienced a major merger since a 
redshift of 1 or 2, then we find $0.41<\sgl<0.45$, closer to the 
observed value.  Thus, perhaps the smaller-than-predicted spread in 
observed disk sizes may be partially understood as the result of the 
special (quiescent) merger history of dark matter halos that harbor 
late-type galaxies. The distribution could be further narrowed by 
assuming that halos with very small spins also form early-type 
galaxies or spheroids via instability processes \citep{mmw:98,bosch:98}. 
 
A second consequence of connecting Hubble type with merger history in 
this way is that the mean value of $\lp$ for the galaxies with a 
quiescent history (no recent major merger) is predicted to be 
$10-15\%$ lower then the mean of all halos.  This implies a similar 
reduction in the prediction for the mean value of the disk scale 
length at a given luminosity.  Whether such a reduction is problematic 
for the model will depend on the detailed arguments used to connect 
disk size to $\lambda$, and will be explored in a future work. 
 
\acknowledgments  
This research has been supported by the Israel Science Foundation 
grant grant 546/98 and by the US-Israel Binational Science Foundation 
grant 98-00217.  AHM acknowledges the support of a Golda Meir 
Fellowship.  We thank James Bullock, Tsafrir Kolatt, Anatoly Klypin, 
Andrey Kravtsov, Cristiano Porciani, Joel Primack and Risa Wechsler 
for stimulating discussions and help with merger trees. 
 
%%%%%%%%%%%%%%%%%%%%%%%%%%%%%%%%%%%%%%%%%%%%%%%% 
\bibliographystyle{apj}         
 
\bibliography{t,spin} 
 
\end{document}